\documentclass[12pt]{article}
\usepackage{epsfig}
\begin{document}
\baselineskip=16pt
\begin{titlepage}
\setcounter{page}{0}

\begin{flushright}
astro-ph/0404225
\end{flushright}

\begin{center}
\vspace{0.5cm}
 {\Large \bf Attractor Behavior of Phantom Cosmology}\\
\vspace{10mm}
Zong-Kuan Guo\footnote{e-mail address: guozk@itp.ac.cn}$^{b}$,
Yun-Song Piao$^{c}$
and
Yuan-Zhong Zhang$^{a,b}$ \\
\vspace{6mm} {\footnotesize{\it
  $^a$CCAST (World Lab.), P.O. Box 8730, Beijing 100080\\
  $^b$Institute of Theoretical Physics, Chinese Academy of Sciences,
      P.O. Box 2735, Beijing 100080, China\\
  $^c$Interdisciplinary Center of Theoretical Studies, Chinese Academy
      of Sciences, P.O. Box 2735, Beijing 100080, China\\}}

\vspace*{5mm} \normalsize
\smallskip
\medskip
\smallskip
\end{center}
\vskip0.6in
\centerline{\large\bf Abstract}
{We investigate the cosmological attractor of the minimally
coupled, self-interacting phantom field with a positive energy density
but negative pressure. It is proved that the phantom cosmology is rigid
in the sense that there exists a unique attractor solution. We plot the
trajectories in the phase space numerically for the phantom field with three
typical potentials. Phase portraits indicate that an initial kinetic term decays
rapidly and the trajectories reach the unique attractor curve. We find that
the curve corresponds to the slow-climb solution.}

\vspace*{2mm}


\end{titlepage}


A scalar field with negative kinetic energy is proposed to explain the
accelerated expansion of present universe as dark energy~\cite{RRC}.
This form of dark energy with the state equation parameter $w<-1$,
dubbed phantom energy violates the dominant-energy condition.
It was shown that this model is consistent with both recent
observations and classical tests of cosmology, in some cases providing
a better fit than the more familiar models with $w>-1$.
In spite of the fact that the field theory of phantom fields encounters
the problem of stability which one could try to bypass by assuming
them to be effective fields~\cite{CHT,GWG}, it is nevertheless interesting
to study their cosmological implication. Recently, there are many relevant
studies of phantom energy~\cite{SW} and the primordial perturbation
spectrum from various phantom inflation models~\cite{PIA}.

The physical background for phantom type of matter with strongly
negative pressure may be looked for in string theory~\cite{MGK}.
Phantom field may also arise from a bulk viscous stress due to particle
production~\cite{JDB} or in higher-order theories of gravity~\cite{MDP},
Brans-Dicke and non-minimally coupled scalar field theories~\cite{DFT}.
The cosmological models which allow for phantom matter appear naturally
in the mirage cosmology of the braneworld scenario~\cite{KK} and in
k-essence models~\cite{COY}.

In this letter we study the attractor behavior of phantom cosmology.
Using the Hamilton-Jacobi formalism, we prove that there exists a unique
attractor solution in the early universe containing a minimally coupled,
self-interacting phantom field. We use an explicit numerical computation
of the phase space trajectories for the phantom field with three typical
potentials. Phase portraits indicate that the initial kinetic term decays
rapidly and the trajectories reach the unique attractor curve. The attractor
curve corresponds to the slow-climb solution.


The action of the phantom field minimally coupled to gravity can be
written as
\begin{equation}
S=\int d^4x\sqrt{-g}\left(\frac{1}{2\kappa ^2}R
 -\frac{1}{2}g^{\mu \nu}\partial _\mu \phi \partial _\nu \phi
 +V(\phi)\right),
\end{equation}
where $\kappa^2 \equiv 8\pi G_N$ is the gravitational coupling, $V(\phi)$
is the phantom potential and the metric signature $(-,+,+,+)$ is employed.
The climbing phantom in a spatially flat FRW cosmological model can be
described by a fluid with a positive energy density $\rho$ and a negative
pressure $P$ given by
\begin{eqnarray}
\rho &=& -\frac{1}{2}\dot{\phi}^2+V(\phi), \\
P &=& -\frac{1}{2}\dot{\phi}^2-V(\phi),
\end{eqnarray}
which means that phantom energy violates the dominant energy condition,
$\rho+P<0$ and $\rho>0$.
The corresponding equation of state parameter is now given by
\begin{equation}
\label{ES}
w \equiv \frac{P}{\rho}
 =\frac{\frac{1}{2}\dot{\phi}^2+V(\phi)}{\frac{1}{2}\dot{\phi}^2-V(\phi)}.
\end{equation}
Since the phantom energy density is positive, Eq.(\ref{ES}) indicates
that $w<-1$, and $w \to -1$ as the ratio $\dot{\phi}^2/2V(\phi) \to 0$.
The evolution equation of the phantom field and the Friedmann constraint
are
\begin{equation}
\label{EP}
\ddot{\phi}+3H\dot{\phi}-V'(\phi)=0,
\end{equation}
\begin{equation}
\label{EF}
H^2=\frac{\kappa^2}{3}\left[-\frac{1}{2}\dot{\phi}^2+V(\phi)\right].
\end{equation}
Note that the sign of the potential force term is negative in Eq.(\ref{EP}),
which distinguishes the phantom field from the ordinary field and
implies that the phantom field climbs up the potential.

The Hamilton-Jacobi formulation is a powerful way of rewriting the
equations of motion, which allows an easier derivation of many inflation
results. We concentrate here on the homogeneous situation as applied to
spatially flat cosmologies, and demonstrate the stability of the phantom
cosmology using the Hamilton-Jacobi formalism~\cite{LPB}.

Differentiating Eq.(\ref{EF}) with respect to $t$ and substituting in
Eq.(\ref{EP}) gives
\begin{equation}
\label{HJ1}
\dot{\phi}=\frac{2}{\kappa^2}H^\prime(\phi).
\end{equation}
This allows us to write the Friedmann equation in the first-order form
\begin{equation}
\label{HJ2}
[H^\prime(\phi)]^2+\frac{3}{2}\kappa^2 H^2(\phi)=\frac{\kappa^4}{2}V(\phi).
\end{equation}
Eqs.(\ref{HJ1}) and (\ref{HJ2}) are the Hamilton-Jacobi equations. They
allow us to consider $H(\phi)$, rather than $V(\phi)$, as the fundamental
quantity to be specified.


Suppose $H(\phi,p)$ is the solution of Eq.(\ref{HJ2}) which is uniquely
determined once the initial conditions have been specified, where the
parameter $p$ is associated with each solution. The general solution to
Eq.(\ref{HJ1}) can be expressed as
\begin{equation}
\label{GS}
a(\phi,p)=a_i \exp \left[\frac{\kappa^2}{2}\int_{\phi_i}^{\phi}d\phi H(\phi,p)
\left(\frac{\partial H(\phi,p)}{\partial\phi}\right)^{-1}\right],
\end{equation}
where $a_i$ is the value at some initial point $\phi_i$.
We consider two solutions $H(\phi,p+\Delta p)$ and $H(\phi,p)$, where
$\Delta p \ll 1$. By differentiating Eq.(\ref{HJ2}) with respect to $p$
and using Eq.(\ref{GS}), we obtain
\begin{equation}
\label{HD}
H(\phi,p+\Delta p)-H(\phi,p) \propto a^{-3}(\phi,p)\Delta p.
\end{equation}
We find that any two solutions approach the attractor solution $H(\phi)$
in an expanding universe. However, the attractor may not be the same
for all values of the parameters. In order to check whether there exists
a unique attractor, one defines the quantity~\cite{LID}
\begin{equation}
F \equiv \left|\frac{H(\phi,p+\Delta p)}{H(\phi,p)}-1\right|.
\end{equation}
Eq.(\ref{HD}) shows that $F \to 0$ as the universe expands, but the form
of the attractor may vary if $\partial F/\partial \phi$ changes sign. Note
that $F$ can go to zero for $\partial F/\partial \phi > 0$ or for
$\partial F/\partial \phi < 0$, but not for both. Thus, if
$\partial F/\partial \phi=0$ for some value of the parameters, the attractor
solution will not be unique for all values of the parameters.
\begin{eqnarray}
\frac{\partial \ln F}{\partial \phi} &=& -H(\phi,p)\left(\frac{\partial H(\phi,p)}
{\partial\phi}\right)^{-1} \nonumber \\
&& \times \left[\frac{3\kappa^2}{2}+\frac{1}{H^2(\phi,p)}\left(
\frac{\partial H(\phi,p)}{\partial\phi}\right)^2\right],
\end{eqnarray}
which indicates that there exists a unique attractor because
$\partial F/\partial \phi$ can not pass through zero.


To study an explicit numerical computation of the phase space trajectories,
it is most convenient to rewrite the evolution
Eqs.(\ref{EP}) and (\ref{EF}) for $H$ and $\phi$ as a set of two
first-order differential equations with two independent variables $\phi$ and
$\dot\phi$
\begin{eqnarray}
\frac{d\phi}{dt} & = & \dot{\phi}, \\
\frac{d\dot{\phi}}{dt} & = & -\sqrt{3}\kappa
\left[-\frac{1}{2}\dot{\phi}^2+V(\phi)\right]^{\frac{1}{2}} \dot{\phi}+V'(\phi).
\end{eqnarray}

Let us consider three typical potentials for the phantom field:
a quadratic potential $V(\phi)=m^2\phi^2/2$,
an exponential potential $V(\phi)=V_0\exp(-\lambda \kappa \phi)$
and a potential of the form $V(\phi)=V_0[1+\cos(\phi/f)]$.
We choose different initial conditions $\phi_0$ and $\dot{\phi}_0$ in the
range $\dot{\phi_0}^2 < 2V(\phi_0)$, and obtain the phase portraits
(Fig.1, Fig.2 and Fig.3, respectively) in the $(\phi,\dot{\phi})$ phase plane.
\begin{figure}
\begin{center}
\includegraphics[angle=-90,scale=0.5]{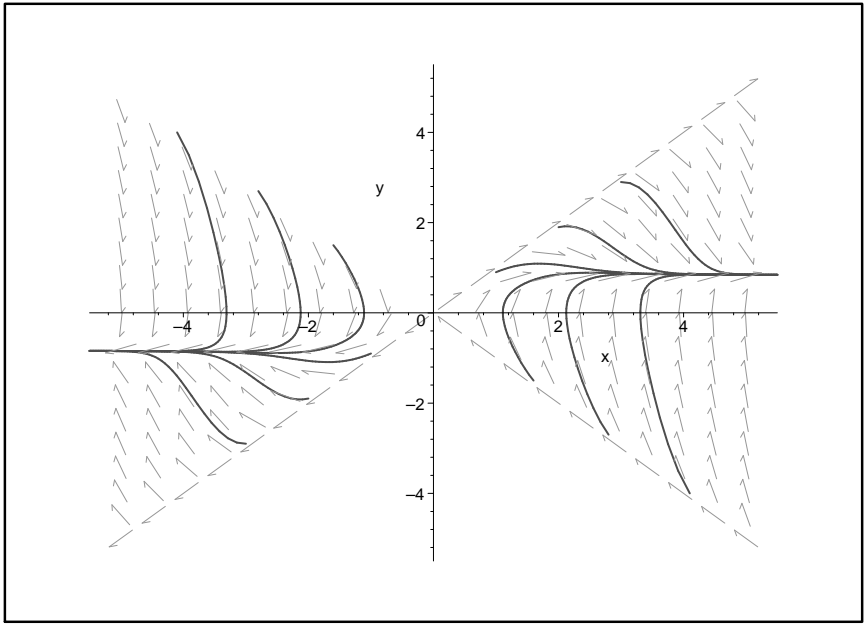}
\caption{Phase portrait for the phantom field with the potential
$V(\phi)=m^2\phi^2/2$ in the $(\phi,\dot{\phi})$ phase plane.}
\end{center}
\end{figure}
\begin{figure}
\begin{center}
\includegraphics[angle=-90,scale=0.5]{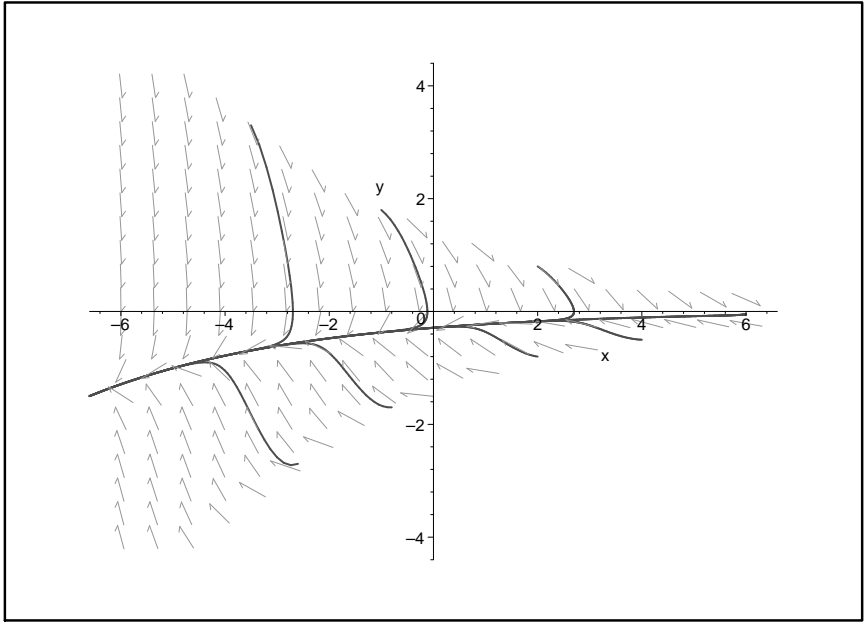}
\caption{Phase portrait for the phantom field with the potential
$V(\phi)=V_0\exp(-\lambda \kappa \phi)$ in the $(\phi,\dot{\phi})$
phase plane.}
\end{center}
\end{figure}
\begin{figure}
\begin{center}
\includegraphics[angle=-90,scale=0.5]{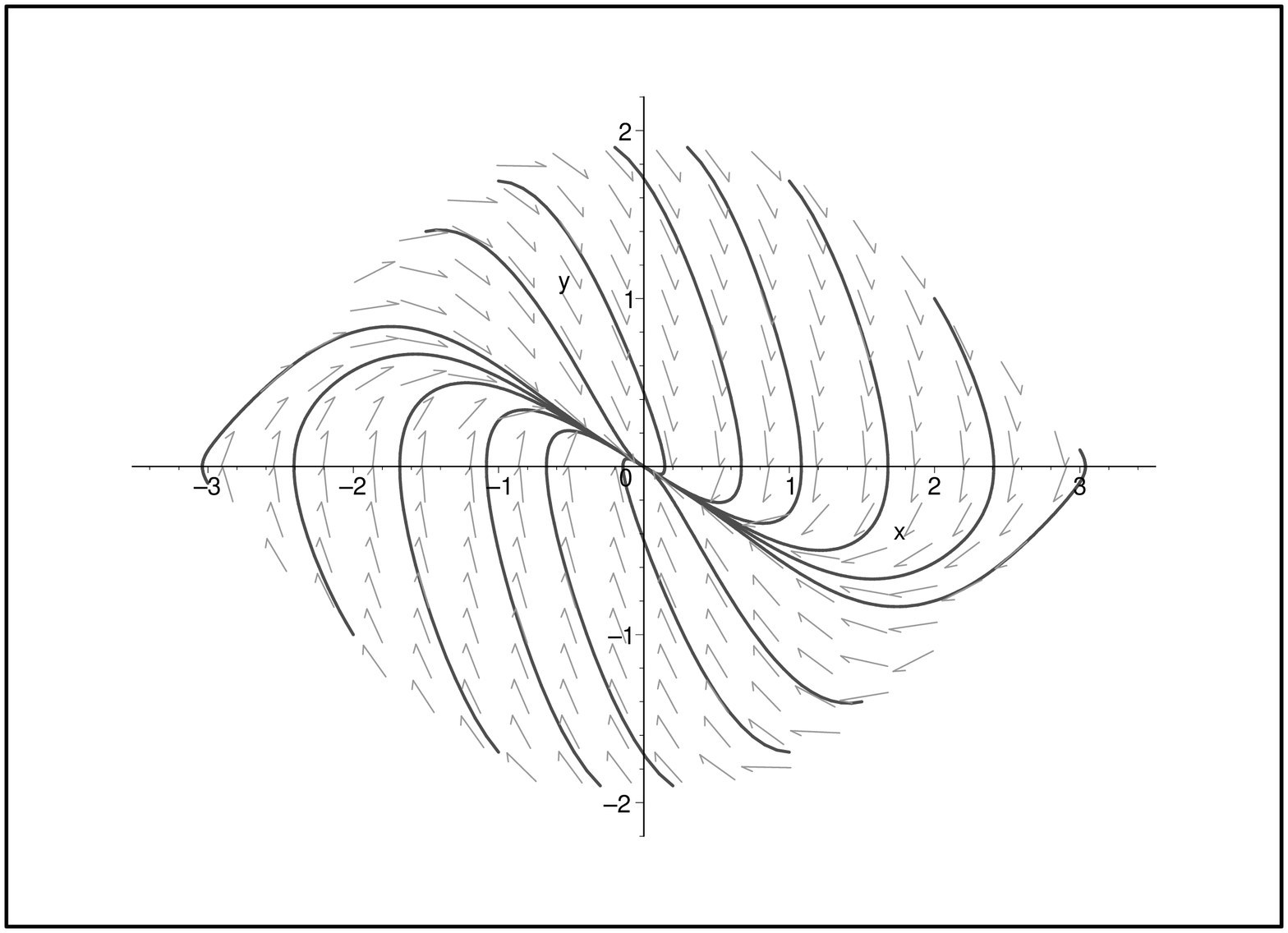}
\caption{Phase portrait for the phantom field with the potential
$V(\phi)=V_0[1+\cos(\phi/f)]$ in the $(\phi,\dot{\phi})$ phase plane.}
\end{center}
\end{figure}

\emph{Power-law potential}. In Figure 1, we see that there is a unique
curve that attracts most of the trajectories in each branch of the two
non-connected regions. An initial kinetic term decays rapidly.
The phantom field climbs up the potential, which
differs from the rolling-down behavior of the normal scalar field.
The two attractor curves correspond to the slow-climb solutions:
\begin{equation}
\dot \phi =\frac{\sqrt{6}m}{3\kappa}
\end{equation}
in the region $\phi > 0$ and $-m \phi < \dot{\phi} < m \phi$, and
\begin{equation}
\dot \phi =-\frac{\sqrt{6}m}{3\kappa}
\end{equation}
in the region $\phi < 0$ and $m \phi < \dot{\phi} < -m \phi$.
In the model, the ratio of kinetic to potential energy of the phantom field
tends to zero, so $w \to -1$.
The general features of the behavior of the trajectories do not change
when we use a power-law potential $V(\phi)=\lambda \phi ^\alpha$.
Although for $\alpha >2$ the kinetic term is no more a constant value
and increases as the universe expands, the ratio of kinetic to potential
energy still tends to zero proportional to $\phi^{-2}$.
For $\alpha =4$ the slow-climb regime becomes the regime of an
exponential growth of the phantom field.

\emph{Exponent potential}. In Figure 2, the attractor curve corresponds to
the slow-climb solution:
\begin{equation}
\dot \phi =-\lambda \sqrt{\frac{V_0}{3}} e^{-\frac{1}{2}\lambda\kappa\phi}
\end{equation}
in the region $-\sqrt{2V_0}\exp(-\lambda \kappa \phi/2) <
\dot{\phi} < \sqrt{2V_0}\exp(-\lambda \kappa \phi/2)$. In the model,
though the kinetic term grows exponentially the ratio of kinetic to
potential energy of the phantom field is a constant value $\lambda^2/6$,
so $w=(\lambda^2+6)/(\lambda^2-6) < -1$.
Therefore, for a phantom field with an exponential potential
$V(\phi)=V_0\exp(-\lambda \kappa \phi)$, the scaling solution exists and
is stable as long as $\lambda^2 < 6$, while for $\lambda^2 > 6$ the
slow-climb approximation breaks down.
This regime is analogous to the scaling solution for a normal scalar field
with an exponential potential~\cite{GPZ}.

\emph{Cosine potential}. In Figure 3, the attractor curve corresponds to
the slow-climb solution:
\begin{equation}
\dot \phi =-\frac{\sqrt{V_0}}{\sqrt{3}\kappa f}
\left(1+\cos\frac{\phi}{f}\right)^{-\frac{1}{2}}\sin\frac{\phi}{f}
\end{equation}
in the region $-\sqrt{2V_0}[1+\cos(\phi/f)]^{1/2} < \dot{\phi}
< \sqrt{2V_0}[1+\cos(\phi/f)]^{1/2}$. The phantom field moves
towards the top of the potential and then settles at $\phi=0$.
The friction term $3H\dot{\phi}$ is so strong compared to the force term
$V'(\phi)$ that the phantom field can not oscillate about the maximum of
the potential, which differs from the regime for a gaussian
potential~\cite{CHT}. In this case, the ratio
of kinetic to potential energy of the phantom field tends to zero,
so the state equation parameter $w$ tends to $-1$. Thus the de-Sitter
like solution is the late-time attractor of the model.


In conclusion, we have investigated the attractor behavior of phantom
cosmology. It is proved that there exists a unique attractor solution, hence
the dynamical system of a single, minimally coupled phantom field is rigid
irrespective of the nature of the potential. Moreover, we consider three
different model: power-law potential, exponential potential and
cosine potential, and plot the trajectories in the phase space numerically.
Phase portraits indicate that an initial kinetic term decays
rapidly and the trajectories reach the unique attractor curve. We find that
the curve corresponds to the slow-climb solution.
The work can be extended to the braneworld scenario.

\section*{Acknowledgements}

It is a pleasure to acknowledge helpful discussions with Rong-Gen Cai.
This project was in part supported by NBRPC2003CB716300.


\begin{thebibliography}{99}
\bibitem{RRC}
 R.R.Caldwell, Phys.Lett. {\bf B545} (2002) 23.
\bibitem{CHT}
 S.M.Carroll, M.Hoffman and M.Trodden, Phys.Rev. {\bf D68} (2003) 023509.
\bibitem{GWG}
 G.W.Gibbons, hep-th/0302199;\\
 S.Nojiri and S.D.Odintsov, Phys.Lett. {\bf B562} (2003) 147;\\
 S.Nojiri and S.D.Odintsov, Phys.Lett. {\bf B565} (2003) 1.
\bibitem{SW}
 A.E.Schulz and M.White, Phys.Rev. {\bf D64} (2001) 043514;\\
 R.R.Caldwell, M.Kamionkowski and N.N.Weinberg, Phys.Rev.Lett. {\bf 91} (2003) 071301;\\
 M.P.Dabrowski, T.Stachowiak and M.Szydlowski, Phys.Rev. {\bf D68} (2003) 103519;\\
 P.Singh, M.Sami and N.Dadhich, Phys.Rev. {\bf D68} (2003) 023522;\\
 X.H.Meng and P.Wang, hep-ph/0311070;\\
 M.Sami and A.Toporensky, gr-qc/0312009.
\bibitem{PIA}
 Y.S.Piao and E.Zhou, Phys.Rev. {\bf D68} (2003) 083515;\\
 Y.S.Piao and Y.Z.Zhang, astro-ph/0401231.
\bibitem{MGK}
 L.Mersini, M.Bastero-Gil and P.Kanti, Phys.Rev. {\bf D64} (2001) 043508;\\
 M.Bastero-Gil, P.H.Frampton and L.Mersini, Phys.Rev. {\bf D65} (2002) 106002;\\
 P.H.Frampton, Phys.Lett. {\bf B555} (2003) 139.
\bibitem{JDB}
 J.D.Barrow, Nucl.Phys. {\bf B310} (1988) 743.
\bibitem{MDP}
 M.D.Pollock, Phys.Lett. {\bf B215} (1988) 635.
\bibitem{DFT}
 D.F.Torres, Phys.Rev. {\bf D66} (2002) 043522.
\bibitem{KK}
 A.Kehagias and E.Kiritsis, JHEP {\bf 9911} (1999) 022.
\bibitem{COY}
 T.Chiba, T.Okabe and M.Yamaguchi, Phys.Rev. {\bf D62} (2000) 023511.
\bibitem{LPB}
 A.R.Liddle, P.Parsons and J.D.Barrow, Phys.Rev. {\bf D50} (1994) 7222;\\
 Z.K.Guo, Y.S.Piao, R.G.Cai and Y.Z.Zhang, Phys.Rev. {\bf D68} (2003) 043508;\\
 Z.K.Guo, H.S.Zhang and Y.Z.Zhang, Phys.Rev. {\bf D69} (2004) 063502.
\bibitem{LID}
 J.E.Lidsey, Gen.Rel.Grav. {\bf 25} (1993) 399;\\
 J.M.Aguirregabiria and R.Lazkoz, gr-qc/0402060.
\bibitem{GPZ}
 Z.K.Guo, Y.S.Piao and Y.Z.Zhang, Phys.Lett. {\bf B568} (2003) 1;\\
 Z.K.Guo, Y.S.Piao, R.G.Cai and Y.Z.Zhang, Phys.Lett. {\bf B576} (2003) 12.
\end{thebibliography}
\end{document}